# All-optical control of the quantum flow of a polariton superfluid

D. Sanvitto[1], S. Pigeon[2], A. Amo[3,4], D. Ballarini[5], M. De Giorgi[1], I. Carusotto[6], R. Hivet[3], F. Pisanello[3], V. G. Sala[3], P. S. Soares-Guimaraes[7], R. Houdré[8], E. Giacobino[3], C. Ciuti[2], A. Bramati[3], G. Gigli[1,5]

[1]NNL, Istituto Nanoscienze - CNR, Via Arnesano, 73100 Lecce, Italy

[2]Laboratoire Matériaux et Phénomènes Quantiques, UMR 7162, Université Paris Diderot-Paris 7 et CNRS, 75013 Paris, France

[3]Laboratoire Kastler Brossel, Université Pierre et Marie Curie-Paris 6, École Normale Supérieure et CNRS, UPMC Case 74, 4 place Jussieu, 75005 Paris, France

[4]CNRS-Laboratoire de Photonique et Nanostructures, Route de Nozay, 91460 Marcoussis, France

[5] Istituto Italiano di Tecnologia, IIT-Lecce, Via Barsanti, 73010 Lecce, Italy

[6] INO-CNR BEC Center and Dipartimento di Fisica, Università di Trento, I-38123 Povo, Italy

[7]Departamento de Física, Universidade Federal de Minas Gerais, Belo Horizonte MG, Brazil

[8]Institut de Physique de la Matière Condensée, Faculté des Sciences de Base, bâtiment de Physique, Station 3, EPFL, CH-1015 Lausanne, Switzerland

**While photons in vacuum are massless particles that do not interact with each other, significant photon-photon interactions appear in suitable nonlinear media, leading to novel hydrodynamic behaviors typical of quantum fluids [1-8]. Here we show the formation of vortex-antivortex pairs in a Bose-Einstein condensate of exciton-polaritons –a coherent gas of strongly dressed photons– flowing at supersonic speed against an artificial potential barrier created and controlled by a light beam in a planar semiconductor microcavity. The observed hydrodynamical phenomenology is in agreement with original theoretical predictions based on the Gross-Pitaevskii equation [9-11,7], recently generalized to the**

**polariton context [12]. However, in contrast to this theoretical work, we show how the initial position and the subsequent trajectory of the vortices crucially depend on the strength and size of the artificial barrier. Additionally, we demonstrate how a suitably tailored optical beam can be used to permanently trap and store the vortices that are hydrodynamically created in the wake of a natural defect. These observations are borne out by time-dependent theoretical simulations based on the formalism of Ref. [12].**

The recent advances in the optical generation, manipulation and diagnostics of quantum gases of exciton-polaritons in planar semiconductor microcavities [13] has paved the way to the study of quantum fluid phenomena in photon gases. Exciton-polaritons are bosonic quasi-particles that emerge in a semiconductor microcavity from the strong coupling of a confined photon and a quantum well exciton [14]. The exciton component provides efficient polariton-polariton interactions which result in strong optical nonlinearities; the photonic component allows for easily generation, manipulation and observation of polariton gases with standard optical techniques. As compared to standard quantum fluids such as liquid Helium or ultracold atomic gases, polariton fluids are expected to show new interesting non-equilibrium effects [15-17] that originate from the finite lifetime of the constituent particles and the driven-dissipative nature of the fluid.

The observation of the phase transition to a Bose-condensed state [18,19] has immediately triggered the search for superfluid behaviours in polariton gases. Superfluidity [20] is manifest in liquid Helium via a number of amazing non-intuitive effects such as the flow without viscosity out of a porous container and the fountain effect. In metals, a superconductive state is visible as permanently running currents for macroscopically long times without any dissipation. In this context, polaritons in planar microcavities open up new exciting possibilities as they provide a unique access and control on all the relevant quantities of the fluid by simple optical means. In particular, one is able to follow in real time the evolution of the density and current profile, as well as of the quantum fluid coherence. This feature is of extreme interest in view of studying hydrodynamic effects in quantum fluids.

Very recently, the physics of polariton propagation has been shown to offer a rich phenomenology related to superfluidity effects. Intriguing effects that exhibited some feature of superfluidity, as expected for a flowing Bose-Einstein condensate, were observed [21]. Following the pioneering proposal of Ref. [22], the occurrence of a superfluid flow in a coherent excitation regime was demonstrated [23]. Persistent currents lasting for macroscopically long times were observed in rotating polariton condensates [24] and more recently, evidence of hydrodynamic dark soliton nucleation was shown [12,25].

The present paper investigates the novel, and so far unexplored, regime of a quantum fluid of polaritons hitting a large obstacle at a supersonic speed, and reports the observation of the peculiar excitations that are created in the fluid. Similar experiments were carried out using a condensate of ultracold atoms in [26,27]. In particular, we have been able to follow the evolution of the fluid propagation and the nucleation of vortices at the surface of a suitably tailored obstacle. We show that the nucleation point and the trajectory of the vortices strongly depend on the shape and height of the barrier as well as on the density of the quantum fluid, demonstrating the possibility to fully control by optical means the vortex appearance and motion. Moreover, using a trapping optical potential, we also show the possibility of permanently storing the vortex pairs that are produced at the surface of a randomly chosen natural defect of the sample. This is a crucial advance with respect to the first experiments showing optical vortices in planar microcavity lasers [28] and in polariton condensates, where pinned vortices spontaneously appeared in the disorder potential of the cavity [29,30] or formed in the minima of the exciting laser field [31].

In the present experiment, polaritons are coherently excited in the planar microcavity by resonantly exciting the sample with a Ti-Sapphire laser. The sample is described in detail in [32]. It consists of a planar DBR microcavity containing 3 $In_{0.04}Ga_{0.96}As$ quantum wells in the cavity layer and a stack of 21 (24) alternated pairs in the top (bottom) mirror. Time resolved real space images of the polariton field are obtained in transmission geometry with the use of a synchroscan streak camera. The real-space phase pattern of the polariton field is inferred from interferograms resulting from the interference of the cavity emission with a reference beam of constant phase coming from the pulsed laser itself. In order for the reference beam to temporally match the whole duration of the polariton dynamics in the cavity, the optical pulse in the reference beam is stretched from a few hundred of femtoseconds to a hundred of picoseconds.

Figure 1 shows the typical lower polariton dispersion of a microcavity under strong photon-exciton coupling. The polariton condensate is excited by a resonant laser with linear vertical polarization and a finite in-plane wavevector as indicated by the red arrow in the figure. The finite in-plane wavevector results in a finite flow velocity of the polariton fluid. We use the excitation scheme proposed in [12]: half of the (otherwise Gaussian) laser spot is masked with a sharp metallic edge so to restrict excitation of polaritons to the upstream region and let them spontaneously flow against the barrier. In this way, the phase of the superfluid is not locked by the excitation laser and can freely evolve in space and time.

The artificial potential barrier is created by a CW laser with opposite linear (horizontal) polarization. This beam is normally incident to the sample and its frequency is resonant with the bottom of the polariton dispersion at $k_{//}=0$. Its purpose is to introduce in the spot region a constant population of polaritons large enough to locally result in a significant blueshift of the polariton states under the effect of the polariton interactions. By varying the CW intensity, blue shifts ranging from 0.1 meV to 0.8 meV are obtained. The size of the potential obstacle is simply controlled by the size of the CW laser spot [33]. Figure 1 shows a scheme of such a potential landscape. In the images shown in Figures 2-4, we isolate the emission from the flowing polaritons injected by the pulsed laser by means of suitable polarizers.

Figure 2 shows five temporal snapshots of the real space density and phase profile of a polariton condensate injected in the cavity with a rightward velocity $v_f$ = 1.1 µm/ps. A rough estimation of the sound speed can be obtained from the aperture angle $\alpha$ of the Čerenkov-like conical density modulation [23] that is visible past the defect in the time-integrated images via the Čerenkov relation [34] $\sin(\alpha/2)=c_s/v_f$. In our case, the polaritons are in fact flowing against the obstacle at a supersonic speed ($c_s<v_f$). The shape of the barrier is Gaussian (~10 µm in diameter, 0.4 meV of mean height) and its position is indicated by a blue circle.

The successive snapshots show the formation of a pair of vortices with opposite circulation in the centre of the gas soon after the polariton flow has met the defect. Vortices are revealed as density minima corresponding to their core in Fig. 2b-e, and as fork-like dislocations in the interferograms shown in Fig. 2g-m. In the first 10 ps the vortex and the antivortex are pushed away from the center of the potential barrier towards its sides at a velocity of ~ 0.9 µm/ps. However, after the vortices have reached the equator of the defect (the equator axis being defined to be perpendicular to the flow direction), there is a clear deceleration of their motion and a small excursion along the vertical direction orthogonal to the flow direction. After most of the polariton pulse has gone past the defect, the vortices keep wandering at a distance of a few tenths of microns from the defect, their motion being mostly oriented in the orthogonal direction with respect to the original injected flow. It is important to remind that the data plotted in the figure are the result of the integration of several billions of nominally identical single realizations: the fact that vortices are visible in the averaged images proves that, in contrast to the purely CW experiment [35], they are formed at the same position and follow the same trajectory at each shot. The good visibility of the fringes in the interference pattern shows that the polariton fluid preserves its coherence during the whole evolution.

Quite surprisingly, the formation of vortices takes place at a position right upstream of the defect and not in its wake as it was instead predicted by theoretical calculations [9-11]. These works, however considered a stationary flow hitting a small and spatially abrupt potential. In contrast, in the present experiments the optically imprinted potential has a smooth, Gaussian-like profile; furthermore, polaritons are injected by a short pulsed laser and their density swiftly decreases in time after the pulse. It is important to point out here that the observed hydrodynamic nucleation of vortex-antivortex pairs can occur up to more than 5 micrometers before the defect. This is fully confirmed by the solution of the time-dependent nonequilibrium Gross-Pitaevskii equations with the experimental excitation conditions (see Fig. 2 third row). We have found that this effect is primarily due to the finite size of the exciting spot, which injects the flowing polariton fluid. Indeed, a local violation of the Landau criterion (responsible for the hydrodynamical instability leading to vortex-antivortex formation) can occur before the defect because of the lowering of the density and consequently of the corresponding Bogoliubov sound speed. Note that in the case of an excitation which corresponds to half a plane [12] the formation is predicted to occur, instead, around the equator of the defect.

A more detailed analysis of the vortex trajectories for different intensities of CW laser creating the optical potential barrier is shown in Figure 3. Increasing the CW power results in the increase of both the height and the effective width of the potential experienced by the moving polaritons. As a consequence, the vortex formation process takes place further upstream from the center of the barrier, where the density is higher (higher $c_s$). For this reason, nucleation simultaneously occurs further away from the defect axis, at positions with a higher local flow velocity, where $v_f=c_s$. These remarkable features are clearly visible by comparing in Fig. 3g the vortex trajectories for the different CW intensities of Fig. 3b-d, and it is reproduced by the solution of the Gross-Pitaevskii equation for polaritons, as shown in Fig. 3f. Note that for the smallest CW power considered here (see Fig. 3a), the vortex-antivortex pair follows the contour of the obstacle and then recombine right downstream of the defect. In this case, thanks to the small defect size, the polariton density is not strongly reduced in the wake of the potential, allowing for the observation of vortex/antivortex annihilation once the flow has recovered its unperturbed pathway.

A quantitative estimation of the polariton density $|\psi|^2$ at the vortex nucleation points can be done by using the equation $c_s = \sqrt{\hbar g |\psi|^2 / m}$, where $g$ is the polariton-polariton interaction constant, $m$ is the effective mass of the lower polariton branch and $c_s$ is the sound speed. The value of $c_s$ at the point where the flow first meets the defect can be estimated from the Cerenkov angle as explained above. For different values of the CW power equal to 6 mW (Fig. 3b), 26 mW (Fig. 3c) and 52 mW (Fig. 3d), the estimated speed of sound is 0.3 µm/ps, 0.4 µm/ps and 0.5 µm/ps, respectively. The corresponding increase of the polariton density goes from $|\psi|^2$ = 9.5 µm$^{-2}$, to $|\psi|^2$ =12.4 µm$^{-2}$ and $|\psi|^2$ = 15.9 µm$^{-2}$, respectively. This calculation is confirmed by the direct estimation of the density from the experimentally measured intensity of the emitted polaritons, which demonstrates the direct relation between the changes in $c_s$ and the ones in the local density variation due to polariton decay.

To further assess the effect of the change in polariton density on the interaction with the defect, we have repeated the experiment with a higher power of the pulsed laser, so to inject a higher density of polaritons at a given value of the barrier height and size (Figs. 3c and 3e). The higher injected polariton density (39.8 µm$^{-2}$ compared to 12.4 µm$^{-2}$ for lower excitation power) corresponds to an increase of the sound speed from $c_s$ = 0.4 µm/ps (Fig. 3c) to 0.8 µm/ps (Fig. 3e). As a result, the nucleation point shifts laterally towards the equator of the defect where the flow velocity is higher. Even though the increase of the density is slightly compensated by a higher fluid velocity ($v_f$ = 1.4 µm/ps) [35], still the dominating effect is in the variation of the speed of sound, giving an increase of 25% in $c_s/v_f$ respect to the experiment in Fig. 3 d.

So far we have shown that an artificial, optically induced, potential barrier is able to produce an hydrodynamic nucleation of vortices in a flowing polariton fluid. However, these vortices can only last for as long as the polariton fluid survives in the cavity. An alternative strategy to make vortices last for macroscopically long times is to work under a CW excitation and to use a suitably tailored mask to nucleate and then trap vortices, as theoretically suggested in [12].

This proposal is experimentally demonstrated in Figure 4. Polaritons with a well-defined momentum are injected by a single CW laser and sent against a potential barrier (here formed by a natural defect present in the microcavity). In the absence of the mask (Fig. 4a), the phase of the polariton fluid is locked to the one of the pump laser, so that vortices are prevented from nucleating even at supersonic speeds, as shown by the homogeneity of the interferogram in Fig. 4f. On the other hand, when a dark region is placed right downstream of a potential (dark triangle in Fig. 4b), pairs of vortices that are hydrodynamically nucleated in the proximity of the defect, can freely penetrate in the dark area where the polariton phase is not locked by the incident laser (note the phase dislocations in Fig. 4g). As the phase of the fluid is homogeneous outside the triangle, vortices cannot diffuse out and get permanently trapped within the triangle borders for macroscopically long times. Moving the dark region slightly away from the defect, we can still appreciate the presence of some trapped vortices within this area. On the other hand, if the dark trap is moved too far away from the defect (Fig. 4e and 4j) vortex nucleation is frustrated by the homogeneous phase that is imprinted by the pump laser. As a result, no vortex is present in the triangular dark region. This observation demonstrate the possibility of a permanent storage of vortices, ready to be further manipulated.

These experiments show the potential of optical methods for the study of the dynamics of quantized vortices in a quantum fluid. During the preparation of this paper, we got aware of a related study on spontaneous nucleation of vortex-antivortex pairs due to natural defects [36]. In the present work, though, we go significantly beyond: we show that we can fully control, via laser-induced artificial potentials, the obstacle parameters, allowing to inspect the physics of the vortex nucleation process. Moreover, we demonstrate the trapping and storing of vortices by using a triangular optically-induced potential. In particular our results confirm that polariton condensates are an alternative to ultracold atoms [37] for the study of quantum turbulence effects [38] in new regimes, and using solid state samples that are operational up to room temperature [39,40]. The all-optical control and trapping of vortices set the premises to study optical vortex lattices and their excitations.

**Acknowledgements**

We would like to thank the support of the POLATOM ESF Research Networking Program. This work was partially supported by the Agence Nationale pour la Recherche (GEMINI 07NANO 07043), the IFRAF (Institut Francilien pour les atomes froids) and the project MIUR FIRB ItalNanoNet. I.C. acknowledges financial support from ERC through the QGBE grant. A.B. and C.C. are members of Institut Universitaire de France (IUF). We are grateful to G. Martiradonna in helping for the realisation of the laser mask and to P. Cazzato for the technical support.

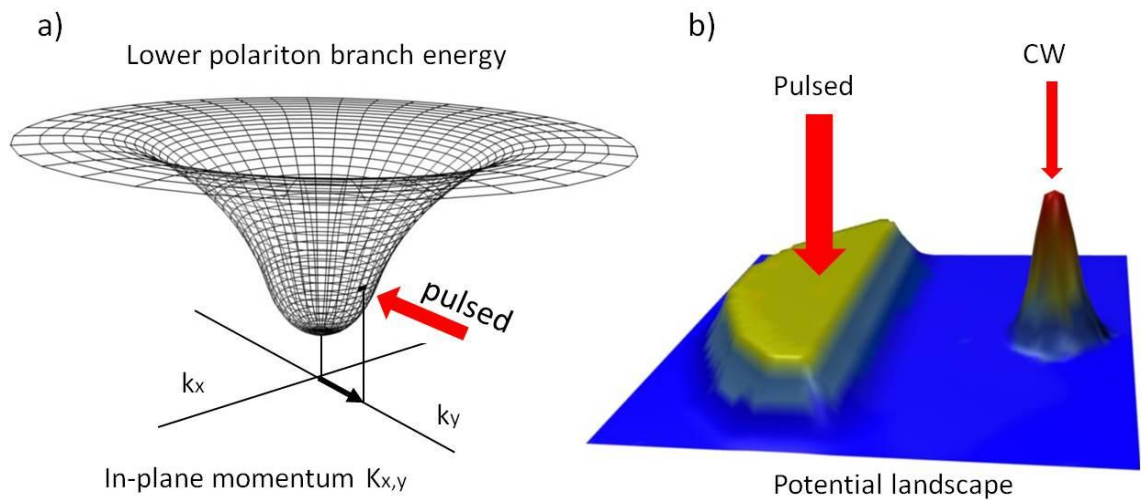

Figure 1. Schematic description of the experiment. (a) lower polariton dispersion with a red arrow indicating the pulsed laser wavevector. (b) Drawings of potential landscape, with the flat potential of the bare polariton state modified by both the pulsed laser injecting the polariton flow and by the CW laser giving origin to the artificial potential barrier.

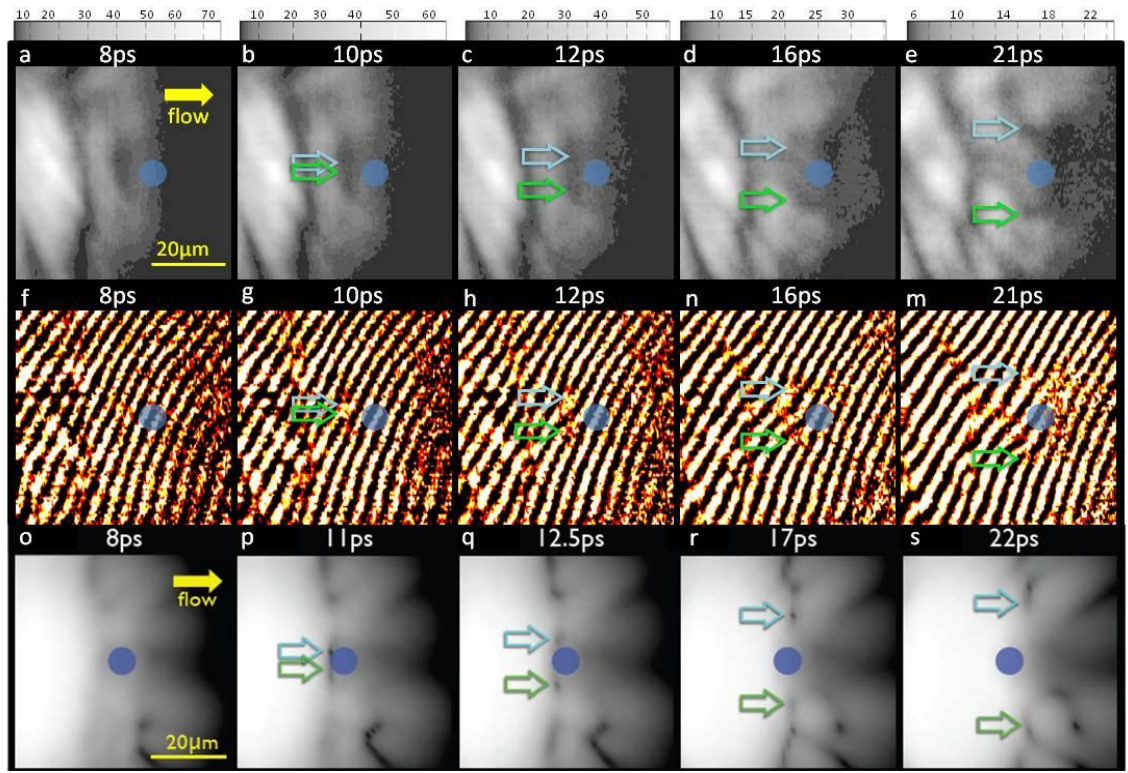

Fig.2: Upper row (a-e): snapshots of the real space emission pattern of the polariton fluid (power of the pulsed laser, $P_{pulsed}$=4mW) hitting an optical defect (blue circle, power of the continuous wave laser, $P_{cw}$=26mW ) at different times (8ps, 10ps, 12ps, 16ps and 21ps after the maximum intensity of the pulse). The dark vertical contour originates from the red-shifted region created by the sharp edge of the masked laser spot. Medium row (f-m): corresponding interferograms giving the spatial profile of the condensate phase. These images are obtained by mixing the real space emission of the condensate with a reference beam of constant phase. Lower row (o-s): theoretical simulations using the parameters of the experimental images (a-e). The agreement in the location of the nucleation process and in the following trajectory is remarkable. The arrows in (b), (g) and (p) indicate the position of the vortex pair nucleation. In (c-e), (h-m) and (q-s) the arrows follow the positions of the vortex-antivortex pair as these are dragged by the polariton fluid on either side of the obstacle. Real space images are in logarithmic color scales.

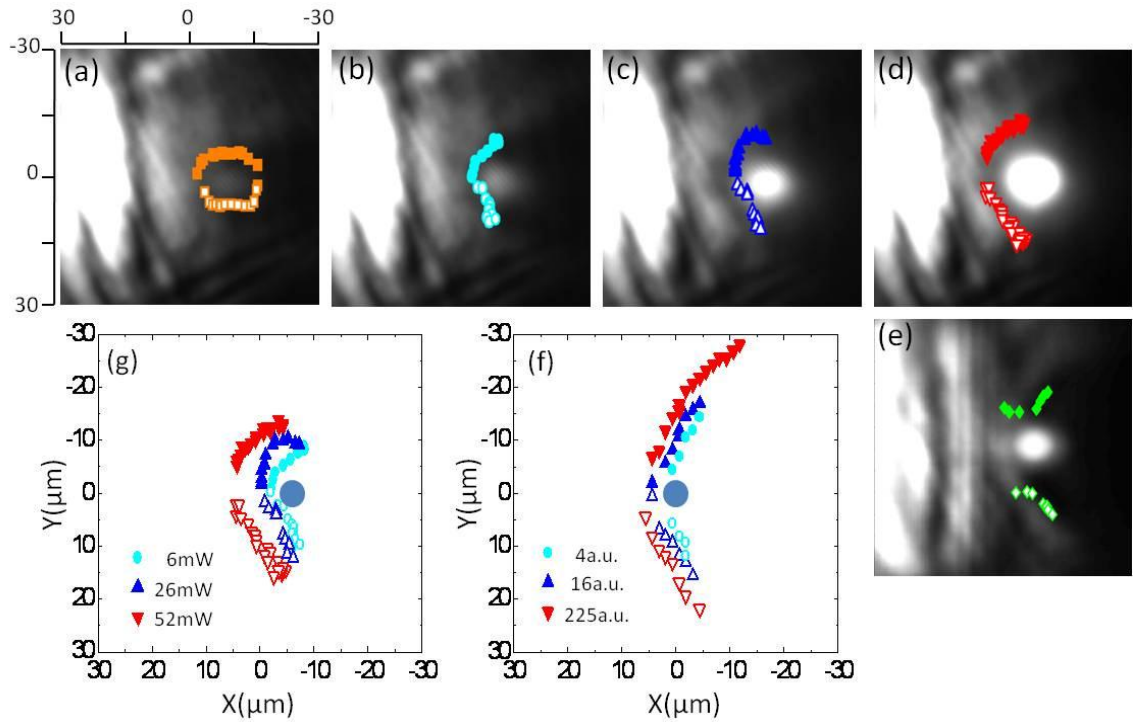

Figure 3. Time-integrated real space emission patterns and corresponding vortex trajectories for different values of the power of the CW and pulsed lasers. Panels (a-d) refer to a fixed pulse power, $P_{pulsed}$=4mW, and different CW powers, $P_{cw}$=2mW, 6mW, 26mW and 52mW respectively. Panel (e) refers to the same potential barrier as in panel (c), $P_{cw}$=26mW, but a higher pump intensity, $P_{pulsed}$=14mW. The time-integrated images in panels (a-e) have been taken with a CCD camera. The bright spot is due to the weak but non-zero transmission of the CW beam through the polarizer and indicates the position and shape of the defect. Panel (g) shows vortex/antivortex trajectories corresponding to images (b), (c) and (d). The trajectories are obtained for each temporal frame by recording the position of the center of the vortex core evidenced via the fork dislocation of the interference pattern (see Fig. 2). Panel (f) displays the theoretical trajectories of different simulation runs using increasing CW pump powers, showing the same trend observed in the experiments. The position of the defect potential is indicated by the blue circle.

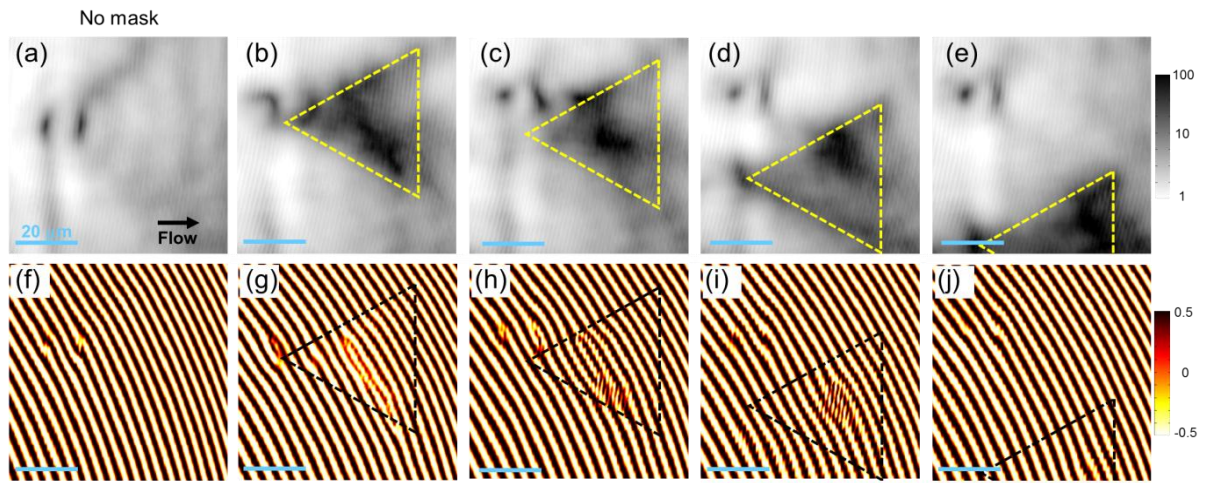

Figure 4: Real space images (a-e) and corresponding interferograms (f-j) for polaritons injected by a CW laser beam and flowing against a natural photonic defect in the microcavity. In (a,f), no mask is applied to the Gaussian spot, so that the density is almost uniform behind the defect and the phase is pinned. The position of the defect corresponds to the black spots in (a). In (b), a dark region is created behind the defect and a pair of vortices nucleates and gets trapped inside the triangular area, as evidenced by the phase dislocations shown in the interferogram (g-i). In (c-e) and (h-j), the triangular mask is laterally shifted with respect to the optical defect. In (c,d,h,i), the lateral shift is small enough for vortices to still nucleate and be visible inside the triangular trap. In (e,j), the mask has been shifted too far away from the defect and vortex nucleation is frustrated due to the homogeneous phase imprinted by the pump laser (j).